\documentclass[universe,article,accept,pdftex,moreauthors]{Definitions/mdpi} 
\firstpage{1} 
\pubvolume{11}
\issuenum{2}
\articlenumber{50}
\pubyear{2025}
\copyrightyear{2025}
\externaleditor{Panayiotis Stavrinos and Øyvind Grøn}
\datereceived{20 November 2024} 
\daterevised{22 January 2025} 
\dateaccepted{25 January 2025} 
\datepublished{5 February 2025} 
\hreflink{https://doi.org/10.3390/\linebreak  universe11020050} 

\usepackage{amsfonts}

\newcommand\ba{\begin{eqnarray}}
\newcommand\ea{\end{eqnarray}}
\newcommand\be{\begin{equation}}
\newcommand\ee{\end{equation}}

\Title{Electromagnetic Waves in Cosmological Space--Time II. Luminosity Distance}

\TitleCitation{Electromagnetic Waves in Cosmological Space--Time II. Luminosity Distance}

\Author{Denitsa Staicova $^{\dagger}$\orcidA{}
 and Michail Stoilov $^{}$*$^{,\dagger}$ }

\Author{Denitsa Staicova$^{1,\S}$\orcidA{} and Michail Stoilov$^{1,*}$}

\AuthorNames{Denitsa Staicova and Michail Stoilov}
\AuthorCitation{Staicova, D.; Stoilov, M.}

\address{
$^{1}$ Bulgarian Academy of Sciences, Institute for Nuclear Research and Nuclear Energy, 72 Tszarigradsko Chaussee, Sofia 1784, Bulgaria\\
$^{\S}$  Email: dstaicova@inrne.bas.bg}

\corres{Correspondence: mstoilov@inrne.bas.bg}

\secondnote{The authors contributed equally to this work.}

\corres{
\hangafter=1 \hangindent=1.05em \hspace{-0.82em} Correspondence: mstoilov@inrne.bas.bg}

\abstract{In this article, we continue our investigation 
 on how the electromagnetic waves propagate in the Friedman--Lema\^itre--Robertson--Walker space{time}. 
Unlike the standard approach, which relies on null geodesics and geometric optics approximation, we derive explicit solutions for electromagnetic waves in expanding spacetime and examine their implications for cosmological observations. 
In particular, our analysis reveals potential modifications to the standard luminosity distance formula.
Its effect on other  cosmological parameters, e.g., the amount of cold dust matter in the Universe, is considered and  estimated from Type Ia supernovae data. 
We see that this alternative model is able to fit the supernova data, but it gives a qualitatively different Universe without a cosmological constant but with stiff or ultra-stiff matter. 
}


\begin{document}



\section{Introduction}
 
The cosmos is probably the first thing that really puzzled humans in the beginning of history.
Its mystery was the origin of the first sciences, and it will probably be one of the last mysteries solved.
For centuries, millions of people gazed into the sky.
For centuries, all our knowledge about the cosmos was obtained from the light of  stars (and planets) that reached our eyes (and later---telescopes) after a long journey through space and time~\mbox{\cite{SS2023, Addazi:2021xuf, LIGOScientific:2017ync, IceCube:2014stg}}. 
Now, we are at the dawn of multi-messenger observations~\cite{Addazi:2021xuf} with gravitational~\cite{LIGOScientific:2017ync} and neutrino~\cite{IceCube:2014stg, Huang:2021hjc} facilities finally providing qualitatively new information.
Nevertheless,  our current understanding of the Universe and its evolution is based on experiments like  Planck~\cite{Planck:2018vyg}, 
Wilkinson Microwave Anisotropy Probe~\cite{WMAP:2012nax}, 
WiggleZ Dark Energy Survey~\cite{Blake:2012pj}, 
Background Imaging of Cosmic Extragalactic Polarization~\cite{BICEP:2021xfz}, 
All Sky Automated Survey for Super Novae~\cite{Neumann:2022szl}, 
Sloan Digital Sky Survey~\cite{Zhao:2021ahg}, 
Hubble Space Telescope~\cite{SupernovaSearchTeam:2004lze,HST:2000azd}, 
James Webb Space Telescope~\cite{Yuan:2022edy},
etc., 
all of them detecting electromagnetic~signals. 

It is obvious that recording celestial electromagnetic signals is not enough.
We need  additional knowledge  in order to correctly interpret  the observed data---we have to find a way to determine cosmic distances across all scales.
Large cosmic distances cannot be measured directly, so we have to rely on theory, which means that distances in cosmology are model-dependent.
In astronomy, distances are determined through the so-called cosmic distance ladder.
This is a succession of methods which allows the estimation of distances, starting from close objects and extending to ones which are arbitrarily far away.
For cosmological distances, the basic approach is to use certain sources as ``standard candles''.
Therefore, knowledge of how light propagates through spacetime is of primordial importance.
The problem is not new and it has been addressed, e.g., in~\cite{Mashhoon:1973obu,Cohen:1974cm, Cabral:2016yxh, Cabral:2016hpq,Asenjo:2017oef} and many others.
One of the possibilities is to treat the electromagnetic waves   according  to the geometric optics approximation.
Using this approach and combining the experimental data with Einstein's theory of gravity, we find that at the moment, the Universe is accelerating its expansion~\cite{Riess1998,Perlmutter1999}; that the flat Friedman--Lema\^itre--Robertson--Walker (FLRW) metric~\cite{P} is most likely the large-scale metric of the Universe beyond, say, 100 Mpc; and that the Universe contains
 approximately 28\% dust matter and 72\% dark energy (cosmological constant).
This is the so-called $\Lambda$CDM model.
 
However, in an expanding Universe, the complete treatment of the electromagnetic wave propagation according to the Einstein--Maxwell equations may lead to a modification of the results from the geometric optics approximation. 
The estimation of distance (especially gravitational distance) is an open question  in alternative gravitational theories as well.
For example, in Ref.~\cite{FFGT}, a quadratic modification of gravity leads to an exponential growth in the ratio of gravity to  electromagnetic luminosity distances.
Other examples of the problem treatment can be found, for instance, in Refs.~\cite{Tomita:1998gf, Sereno:2001cc,  Demianski:2003bb, Fleury2015, Nikolaev:2016imy, Belgacem:2017ihm, LISACosmologyWorkingGroup:2019mwx, Mukherjee:2020mha, Garoffolo:2020vtd,Fonseca:2023uay,  Romano:2023ozy, Gupta:2023mgg, Cuzinatto:2023kbo} and others in connection to gravitational astronomy. 
See also Ref.~\cite{PhysRevD.94.084018} and the references therein.
The reason for such interest is that in the modified gravity theories, the luminosity distance for gravitational waves can differ substantially from that of electromagnetic signals (up to an order of magnitude), suggesting that this effect needs to be examined in detail. 
Therefore, a simultaneous measurement of gravitational and electromagnetic signals from a single cosmic event can shed light on the subject.

In the present work, we try to obtain a formula for the luminosity distance as a result of the explicit functional form of the propagation of electromagnetic waves in the flat FLRW  metric.
Our approach is very close to the one followed in Ref.~\cite{EMM} but without any assumptions on the form of the electromagnetic potential.
We wish to recall that the electromagnetic field in curved spacetime is governed 
not by the simple wave equation $\Box A^\mu = 0$ assumed in the geometric optics approximation, but by an equation with an additional term proportional to the product of the Ricci tensor and the electromagnetic potential.  
The extra term is quite interesting, because for a not-Ricci-flat spacetime,
its effect cannot be nullified even in an inertial coordinate system (and thus, it resembles a tidal force).
For instance, in the case of an exponentially expanding Universe (which happened during the inflation epoch at the beginning of the Universe and is happening again today), this term effectively renders mass to the electromagnetic potential ~\cite{SS2023} (each of the electromagnetic potential component  satisfies a Klein--Gordon equation).
This, in turn, suggests a non-zero-length world line of the light ray,
which surely affects the wave propagation and, as a consequence, the resulting luminosity distance.
%
Therefore, in order to clarify the problem, we construct an exact solution for the electromagnetic potential suitable for describing light propagation (for example, light emitted by a distant star and observed locally) and we use it to determine the luminosity distance. The luminosity distance obtained in the described way differs from the standard one. We test it against the supernovae of type Ia (SN Ia) data. The fit is successful, but instead of a cosmological constant term, it predicts the presence of another type of exotic matter. We discuss the possible origin of such a term and the consequences of such a model.

This paper is organized as follows: 
First, we sketch the derivation of distance luminosity using the properties of  0-world lines.
Second, we construct a solution of the electromagnetic equation describing the light emitted by a star.
Third, we construct the corresponding stress--energy tensor, we find the energy density and energy flow as functions of the distance between observer and emitter, and we determine the corresponding luminosity distance.
Fourth, because the newly obtained formula for the luminosity distance differs from the standard one, we consider how it affects the matter content of the Universe, mainly using data inferred from SN Ia. 

\section{Redshift, Co-Moving Distance, Luminosity Distance---The Null Distance Approach}
Hereafter, we shall use the Cartesian form of the FLRW metric.
In this case, the invariant length is
\be
ds^2=-dt^2 + a(t)^2(dx^2 + dy^2 + dz^2),\label{id1}
\ee
where $a(t)$ is the scale factor, and we work in units where the speed of light $c=1$.

We begin with a derivation of the redshift and luminosity distance based simply on the 0-world-line assumption.
Consider a signal, propagating in a straight direction with speed $1$ (i.e., $c$), which is emitted at time $t_e$  and is observed at time $t_o$.
%
Without any loss of generality, we can place the origin of the spatial coordinate system at the emission point, and we can consider that the signal is along the $x$-axis.
Therefore, the 0-length world line of the signal is parametrically determined by the equation
\be \{t, x(t), y(t), z(t)\}= \{t, \int_{t_e}^t \frac{d\tau}{a(\tau)}, 0, 0\}. \label{nwl}\ee

Let us denote by $d$ the co-moving distance between observation and emission.
Consider another signal emitted  from the origin, again in the $x$ direction, but at time $t_e+d t_e$. 
It will arrive at $x=d$ at $t_o + d t_o$, where
\be \int_{t_e+d t_e}^{t_o+d t_o} \frac{d\tau}{a(\tau)} = d = 
\int_{t_e}^{t_0} \frac{d\tau}{a(\tau)},\label{cmd0}\ee
so that
\be \frac{d t_o}{a(t_o)} = \frac{d t_e}{a(t_e)}.\label{dt}\ee
If the signal is periodic, then using Equation~(\ref{dt}), we obtain that the redshift at $t_o$ is
\be z=\frac{\lambda_o-\lambda_e}{\lambda_e}=\frac{a(t_o)}{a(t_e)}-1.\ee
If we normalize $a(t)$, so that $a(t_o)=1$, then the equation can be put in the form
\be a=\frac{1}{1+z}. \label{rs}\ee
Taking the derivative of the above equation, we obtain $\dot{z} = -\dot{a}/a^2$, where we denote by dot over the function its time derivative ($\dot{f}\equiv d f/d t, \forall f$).
This formula  can be used to change the variables in the definition of the co-moving distance passed by the signal, so that
\be d = \int_0^z \frac{d\zeta}{(\dot{a}/a)}=
d_H\int_0^z \frac{d\zeta}{\sqrt{\Omega_m (1+\zeta)^3+\Omega_\Lambda}},\label{cmd}
\ee
where we used the Friedman equation $\dot{a}/a\sim\sqrt{\rho}$ to obtain the final equation (for simplicity, we stuck to $\Lambda$CDM model).
In the above equation, $d_H$ is Hubble distance ($d_H=1/H_0$; recall that here the speed of light is $1$), and $\Omega_m$ and $\Omega_\Lambda$ are, respectively, the normalized values of dust matter and dark energy densities at $t_o$.
Note that the integral on  the right hand side  can be evaluated in a closed form (see 
the paragraph after Equation~(\ref{sf})).

A basic quantity called  {\it the luminosity distance} connects the observed energy flux at a co-moving distance of $d$ to the intrinsic luminosity of the isotropic emitter. 
\be S=\frac{(dn/dt_o) h \nu_o}{4\pi d^2}=\frac{(dn/dt_e) h \nu_e}{4\pi d_L^2}.\label{fl}\ee
Here, $d n/d t$ is the total photon flux, $h \nu$ is the energy of a photon with frequency $\nu$, and for simplicity, we consider only monochromatic photons.
It follows  from Equations~(\ref{dt}), (\ref{cmd}), and~(\ref{fl})~that
\be d_L=(1+z) d.\label{ld}\ee

\section{Electromagnetic Waves in FLRW Metric}

In this section, we wish to obtain a formula for the luminosity distance  using exact solutions for electromagnetic wave propagation in FLRW spacetime, thus providing a more fundamental approach. 

The Lagrangian of the free electromagnetic field is 
\be 
\mathfrak{L}=-\frac{1}{4} F_{\mu\nu} F^{\mu\nu}. 
\ee 
Here, $F$ is the electromagnetic field tensor,
$
F_{\mu\nu}= \nabla_\mu A_\nu-\nabla_\nu A_\mu=
\partial_\mu A_\nu-\partial_\nu A_\mu ,
$
and $A_\mu$ is the electromagnetic potential. 
Note that because of the antisymmetric structure of the electromagnetic field tensor, both $F$ and $\mathfrak{L}$ are exactly the same as in the usual Minkowski space{time}. 
The corresponding equation of motion is 
$
\partial_\nu(\sqrt{-g}F^{\mu\nu})= 0,
$
where $g$ is the determinant of the metric tensor $ g_{\mu\nu}$.
This equation can be recast into an explicitly covariant form, namely 
\be \nabla_\nu F^{\mu\nu}= 0\label{em0}.\ee
Here,  $\nabla^\mu$ is the covariant derivative in the space{time}.

 The  electromagnetic field tensor is invariant under the usual gauge symmetry 
$A_\mu(x) \rightarrow A_\mu(x)+\partial_\mu f(x)$, where $f(x)$ is an arbitrary function, and therefore, a gauge fixing is required. 
In this article, we use the generalized Lorenz gauge
\be\nabla_\nu A^\nu=0\label{gauge}\ee
in which  Equation  (\ref{em0}) takes the form
\begin{equation}        
\Box A^{\mu}-R^{\mu}{_\nu}A^{\nu} = 0.  
	\label{EMprop}
\end{equation}
Here,  $\Box=\nabla^\nu \nabla_\nu$ is the d'Alembert operator, and $R^{\mu}_{\nu}$ is the Ricci tensor
$
R_{\mu\nu}= \partial_\alpha \Gamma ^\alpha_{\mu\nu}-\partial_\mu \Gamma ^\alpha_{\alpha \nu}+\Gamma ^\alpha_{\mu\nu}\Gamma^\beta_{\alpha\beta}-\Gamma^\beta_{\alpha\mu}\Gamma^\alpha_{\beta\nu}
$.

Obviously, Equation~(\ref{EMprop}) is not the wave equation  
$\Box A^\mu =0$.
Note that if the space{time} is not Ricci-flat, then the equation satisfied by the electromagnetic waves will differ from the one in the Minkowski space{time} even in the locally inertial coordinate frame.
The plane wave solution of Equation~(\ref{EMprop}) can be found in Ref.~\cite{SS2023}.
There, we have constructed a global solution in radial coordinate system  as well.
At present, we consider this solution as describing dipole radiation, and therefore, it will not be used here.

At this point, we wish to make a short detour  to the geometric optics approach~\cite{Sachs:1961zz, Ehlers:1992dau} to the light propagation in curved spacetime.
This approach relies on the  approximation that the tail term in Equation~(\ref{ellis}) is much smaller than the other one.  This is supposed to be true for small spacetime curvature (which, in turn, is  determined  by the matter fields  dominating the Universe). 
The electromagnetic waves are treated in this approximation  as rays following null geodesics. According to Ref.~\cite{EMM}, in the geometric optics approach, the solution for the electromagnetic potential is supposed to be in the form:
\begin{equation}
A_a = g(\psi)\alpha_a + \text{small tail terms},\label{ellis}
\end{equation}
where $g(\psi)$ is an {\it arbitrary} function of the phase $\psi$, and $\alpha_a$ is an amplitude which is supposed to vary slowly compared to $g$.
Further, based on the arbitrariness of function $g$, it is assumed that all coefficients in front of the derivatives of $g$ of different orders in \mbox{Equations~(\ref{gauge}) and (\ref{EMprop})} are {\it separately} equal to zero, thus giving rise to the following system: 
\begin{align}
2k^b\nabla_b\alpha_a +\alpha_a\nabla_b k^b &=0  \\
\nabla_b\nabla^b\alpha_a + R_{ab}\alpha^b &= 0\\
\alpha^ak_a &= 0,\\
\quad \nabla_a\alpha^a &= 0,\\
\quad k_ak^a &= 0, 
\end{align}
where $k_a = \psi_{,a}$ is the propagation vector.

Under the geometric optics approximation, the observed flux is 
$\mathfrak{F} = T_{ab}u^au^b \approx \alpha^2(k_au^a)^2g(\psi)$, where $\alpha^2$ varies inversely with the cross-sectional area $\delta S$ of the ray bundle. Accounting for the redshift relation $(k_au^a)_o = (k_au^a)_e/(1+z)$, this yields $\mathfrak{F}\delta S = C\frac{d\Omega}{(1+z)^2}$, where $C$ is constant along null geodesics. Since the source luminosity $L = \int_S (1+z)^2\mathfrak{F} dS$ gives $C=\frac{L}{4\pi}$, we obtain the fundamental relation $\mathfrak{F} = \frac{L}{4\pi d_L^2}$ with the $(1+z)^2$ factor emerging naturally from the wave propagation in curved spacetime rather than from the photon energy and count arguments.

We consider the assumptions of  the geometric optics  as too restrictive, 
and in what follows, we shall construct an appropriate solution to Equation~(\ref{EMprop}) without relying on them.
We can further use such solutions to determine the luminosity of an isotropic light source, following the procedure outlined in Ref.~\cite{EMM}.

We are looking for a solution in the form $A^\mu=\mathcal{A}^\mu(t,r)/r$, 
where $r=\sqrt{x^2+y^2+z^2}$. This ansatz naturally incorporates the $1/r$ falloff expected for radiative solutions; it maintains consistency with the spherical symmetry of FLRW spacetime and allows for the explicit tracking of the cosmological expansion effect. 

There is a topological  obstacle (hairy ball theorem) in the construction a global solution describing the electromagnetic potential of spherical waves. 
This is the reason why we construct only a local one, but this is good enough for our purposes, since in cosmology, we can perform only local observations.
%
%
%
We consider a specially orientated coordinate system, where the observer is on the 
$x$-axis, and in its vicinity, we have the following equation for the physical (transverse) components of $\mathcal{A}$:
\be
-\ddot{\mathcal{A}}^i +
\frac{1}{a^2}{\mathcal{A}^i}'' -
5\frac{\dot{a}}{a} \dot{ \mathcal{A}}^i
-2 \left( 2 \frac{\dot{a}^2}{a^2}+\frac{\ddot{a}}{a}\right)\mathcal{A}^i
=0,\;\; i=2, 3. \label{teq}
\ee
Here (again), $ \dot{ \mathcal{A}}^i=\partial_t \mathcal{A}^i$ and 
$ {\mathcal{A}^i}'=\partial_x \mathcal{A}^i$.
Surprisingly, Equation~(\ref{teq}) is exactly the equation for a plane wave we found in Ref.~\cite{SS2023}.
Therefore, on the world line $\{t,x,y,z\}=\{t,x,0,0\}$, we have the following solution for the transverse components of the \mbox{electromagnetic potential:}
\be A^i(t,x)=\frac{A^i_\pm}{x\; a(t)^2}
\cos\left(k \int^t \frac{d\tau}{a(\tau)} \pm k x + \phi_\pm\right).\label{sol}
\ee
Here, 
$A^i_\pm$ and $\phi_\pm$ are some constants, and $k$ is another constant indicating the wave number.
In what follows, we shall consider only the outgoing solution (``-'' sign).
Mathematically, our solution satisfies the outgoing boundary condition at infinity.
Due to the singularity at the spatial coordinate origin of the solutions given by Equation~(\ref{sol}), we need some boundary condition at $x>0$ to specify the amplitudes $A^i_-$.
We normalize the solution to match the luminosity of SN Ia at 10 pc.

Using Equation~(\ref{sol}), we can show that the redshift and co-moving distance are given by Equation~(\ref{rs}) and Equations~(\ref{cmd0}) and (\ref{cmd}), respectively.

\section{Luminosity Distance
Based on Equation~(\ref{sol})
}

The energy flux of an electromagnetic wave can be found from its energy density or Poynting vector at the observation point.
For this purpose, we plug the solution found in the previous section into the electromagnetic stress--energy tensor.
In order to work with invariant quantities, we shall need two additional normalized 4-vectors $u$ and $v$, the first one being the 4-velocity of the observer and the second one determining the spatial orientation of the surface through which the electromagnetic energy passes.
Taking into account that $g^{\mu\nu}=\mathrm{diag}\{-1,1/a(t)^2,1/a(t)^2,1/a(t)^2\}$, we define the vectors $u$ and $v$ \mbox{as follows:}
\ba 
u_\mu &=& \{1,0,0,0\},\;\;|u|^2=-1,\\
v_\mu &=& \{0, a(t), 0, 0 \},\;\;|v|^2=1.
\ea 

The stress--energy tensor of the electromagnetic field is 
\be
T_{\mu\nu}=F_{\mu\alpha}F_\nu^{\;\;\alpha}-\frac{1}{4}g_{\mu\nu}F_{\alpha\beta}F^{\alpha\beta}.\label{set} 
\ee
Therefore, the averaged observed energy density at point $\{t,x,0,0\}$ for the solution given by Equation~(\ref{sol}) is
\be
\overline{u\cdot T\cdot u}=\frac{k^2}{2 x^2 a(t)^4}((A_-^2)^2+(A_-^3)^2),\label{ed}
\ee
where we have neglected terms $\sim x^{-3}$ and $\sim x^{-4}$.

The energy flux in the $v$ direction in the observer frame is 
\be
\overline{u\cdot T\cdot v}=\frac{k^2}{2 x^2 a(t)^4}((A_-^2)^2+(A_-^3)^2),\label{ef}
\ee
where, again, we have neglected terms $\sim x^{-3}$ and $\sim x^{-4}$.
Both quantities are the same, as expected; see, e.g.,~\cite{EMM}.

Comparing the flux determined by Equations~(\ref{ed}) and (\ref{ef}) at different distances, \mbox{we obtain}
\be d_L=(1+z)^2 d. \label{nld}\ee

\section{Numerical Fits}

The data for SN Ia have been used in the last decades to determine the geometry of FLRW.
Now, there are really good compilations of SN Ia observations which provide the necessary information. For example, the SCP ``Union2.1'' SN Ia compilation~\cite{union} contains the data for 580 SN Ia
\endnote[1]{Available online: \url{https://supernova.lbl.gov/Union/figures/SCPUnion2.1_mu_vs_z.txt}  (accessed on 27 January 2025). }.
For each  supernova, it contains the corresponding distance modulus and redshift.
The distance modulus is (by definition)
$ d_m=5 \log_{10}(d_L/d_0) $ where \mbox{$d_0=10$ pc,} and $d_L$ is the luminosity distance, given either by Equation~(\ref{ld}) or by Equation~(\ref{nld}).

In the $\Lambda$CDM case, the SN Ia fit is based on  the function
\be f_{\alpha,\beta}(z) =\alpha+ 5 \log_{10}\left((1+z)
\int^{1+z}_1 \frac{d \zeta}{\sqrt{\zeta^3 +\beta}}\right),\label{sf}
\ee
where $\alpha$ and $\beta$ are the fitting parameters. 
Their physical interpretation is $\beta=\Omega_\Lambda/\Omega_m$
and $\alpha=5 \log_{10} (d_H/d_o \sqrt{\Omega_m})$.
The integral in the above equation represents the co-moving distance $d$ (see Equation~(\ref{cmd})) without the multiplicative constant $d_H/\sqrt{\Omega_m}$ which is moved in $\alpha$.
The integral can be taken in a closed form by means of an integral representation of the ordinary hypergeometric function $_2\!F_1[\frac{1}{6},\frac{1}{2};\frac{7}{6}; -\beta/y^3] = \int dy/2\sqrt{y^3+\beta}$.
The integral also has a closed functional expression for other combinations of a two-component Universe 
(including curvature as well), but in the general case, it can only be calculated numerically.
Taking into account that $\Omega_\Lambda + \Omega_m = 1$,
the physical quantities can be extracted from the fitting parameters as follows:  $\Omega_m= 1/(1-\beta)$, $\Omega_\Lambda =1-\Omega_m$, and the dimensionless Hubble constant is $h=3\times 10^{(40-\alpha)/5}/\sqrt{\Omega_m}$.

Equation~(\ref{nld}) differs from Equation~(\ref{ld}),
and correspondingly, the fitting function for the SN Ia data should be different and will lead to different preferable matter content in the Universe.
Hereafter, we consider a Universe made of two components---a cold dust matter with normalized density fraction $\Omega_m$ and additional term  with density fraction  $\Omega_{ex}$ and
equation of state $w_{ex}$, which we consider as a free parameter.
As usual, the densities satisfy the critical density condition 
$\Omega_m + \Omega_{ex} = 1$.
Let us recall that the equation of state in the perfect fluid analogy for the stress tensor relates the pressure $p$ to the energy density $\epsilon$ (matter density $\rho$ for cold matter).
Usually, the dependence is $p= w \epsilon$ (or $p= w\rho$).
For example, the equation of state for radiation is $w = 1/3$; for pressureless (cold) matter, it is $w = 0$; and for the cosmological constant, $w=-1$. 
In any case, the (energy) density scales as $a(t)^{-3(1+w)}$.
Therefore, for the model in consideration, the co-moving distance is as follows:
\be d = \int_{t_e}^{t_o}\frac{d t}{a(t)}= \int_0^z \frac{d\zeta}{(\dot{a}/a)}=
d_H\int_0^z \frac{d\zeta}{\sqrt{\Omega_m (1+\zeta)^3+\Omega_{ex} (1+\zeta)^{3(w_{ex}+1)}}}.\nonumber
\ee
Having in mind these considerations and Equation~(\ref{nld}), we choose the fitting function for the SN Ia data to be
\be \hat{f}_{\alpha,\beta}(z) = \alpha+  5 \log_{10}\left((1+z)^2
\int^{1+z}_1 \frac{d \zeta}{\sqrt{\zeta^3 +\beta \zeta^{3(1+w_{ex})}}}\right).\label{nsf}
\ee
%
%

Note that $\alpha=5 \log_{10} (d_H/d_o \sqrt{\Omega_m})$ in both Equations~(\ref{sf}) and (\ref{nsf}).
Also, $\Omega_m$  is the same: $\Omega_m= 1/(1-\beta)$.
So, Equation~(\ref{nsf}) generalizes Equation~(\ref{sf}),
replacing the cosmological constant with a matter characterized by a free parameter $w_{ex}$.
This matter can, in principle, represent the cosmological constant as well.

\begin{figure}[H]
 \includegraphics[width=0.8\textwidth]{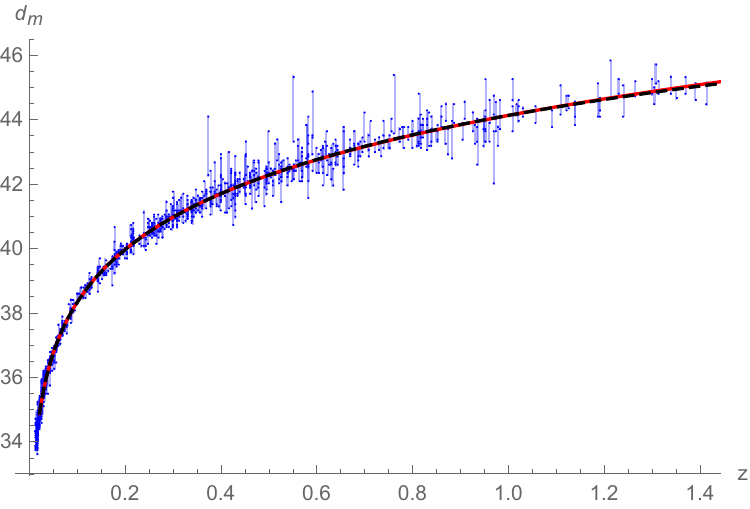}
 \caption{SN Ia fits. Black dashed line---the standard fit based on Equation~(\ref{sf}) with 28\% dust with equation of state $w=0$ and 72 \% dark energy with equation of state $w=-1$. Red line---the new fit, based on Equation~(\ref{nsf}) and a mixture of 70\% dust with equation of state $w=0$ and 30\% matter with equation of state $w=5/3$.
 The fits with other $w_{ex}\ge 1$ differ marginally from it.}
 \label{fig1}
 \end{figure}

\begin{figure}[H]
 \includegraphics[width=0.8\textwidth]{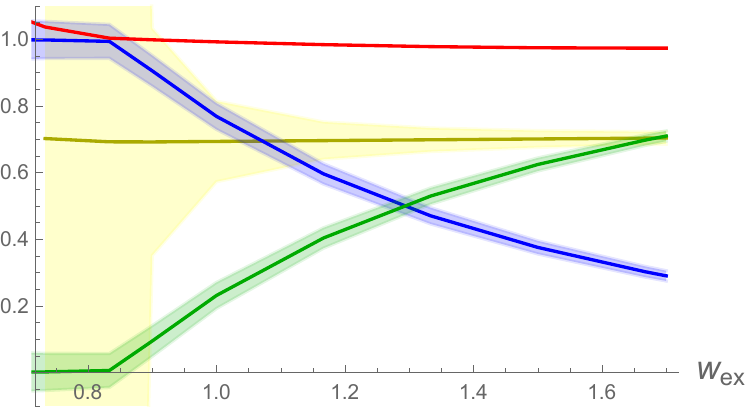}
 \caption{The Universe parameters as functions of the new matter equation of state. In blue---$\Omega_{ex}$ with its standard variation as a shaded corridor. In green---$\Omega_m$ with its standard variation. Yellow---the dimensionless Hubble constant $h$ and its standard variation. The red line gives the fit's 
 estimated variance ($\chi^2/(\mathrm{number~of~degrees~of~freedom})$).}
 \label{fig2}
 \end{figure}

 In Figure~\ref{fig1}, we juxtapose the standard fit and the new one for $w_{ex}=5/3$.
 As can be seen, they are almost identical.
 The fits with other $w_{ex}\ge 1$ differ marginally from them but have different matter content.
 The values of $\Omega_m$, $\Omega_{ex}$, and $h$, and the quality of the fit based on Equation  (\ref{nsf}) as functions of $w_{ex}$ are shown in Figure~\ref{fig2}.
We wish to stress that this fit is possible only for equations of state $w_{ex} \lesssim 1.7$.
It is also clear from Figure~\ref{fig2} that the fit is reasonable for $w_{ex} \gtrsim 0.9$.

Furthermore, we investigate the same fit with the nested sampler Polychord~\cite{Handley:2015fda}), applied to the data of the binned Pantheon dataset with their covariance, consisting $40$~supernovae luminosity measurements in the redshift range $z\in (0.01,2.3)$~\cite{Pan-STARRS1:2017jku}. We do not use the newer Pantheon Plus dataset since this study is a proof of concept for which the Pantheon dataset is good enough. In this case, we do not use the fitting function, but we directly use the modified luminosity distance, plugged into the Friedman equations with various combinations of $\Omega_m$ and higher terms $\sim \Omega_{ex}$---sixth and eighth order on $(1+z)$. 

\vspace{-6pt}

\begin{figure}[H]
 \includegraphics[width=0.8\textwidth]{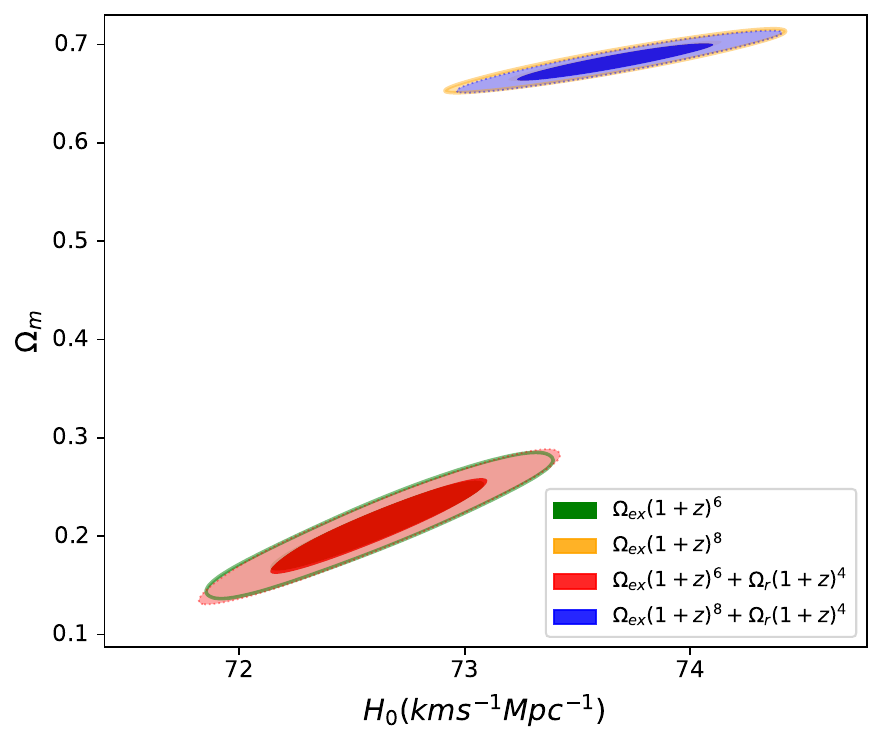}
 \caption{On the plots, the posteriors of the different expansion for the Pantheon dataset. We note two different mean values for $\Omega_m$ corresponding to the 6th and 8th powers terms. Adding radiation does not change the posteriors significantly.}
 \label{MCMC}
 \end{figure}

In Figure~\ref{MCMC}, we show the posteriors of our tests, combining different terms as mentioned in the legend of the plot: $(1+z)^6, (1+z)^8$, with and without radiation term $(1+z)^4$ 
(thus generalizing slightly the model in Equation~(\ref{nsf})). 
Note that we also tested the inclusion of Quasars~\cite{Risaliti:2015zla} to the datasets, and the inferred cosmological parameters were practically the same (up to the second decimal digit); thus, we do not include their contour plot in the figure. 
 
We fix $\Omega_r=0.001$ for simplicity, and because there are numerical problems with the likelihood with a wide uniform prior on $\Omega_r$. 

We obtain that some of the combinations are not convergent, while others are very slowly convergent. On the plot, we show the normally convergent combinations. One can see the contour plots in Figure \ref{MCMC}. Notably, the expansions $\sim (1+z)^6$, ($w_{ex}=1$) are able to fit the expected value of $\Omega_m$ for the standard wide prior ($\Omega\in[0,1]$, while the $(1+z)^8$, ($w_{ex}=5/3$) ones predict a much larger matter density. Both expansions, however, are able to fit the expected $H_0$ in the uniform prior we used, $H_0\in [50, 100]$. Note that the $(1+z)^8$ expansions predict a slightly higher value for $H_0$. Throughout, we have assumed that $\sum(\Omega_i)=1$. Also, the tight constraint on $H_0$, despite our wide prior, comes from fixing $M_B = -19.25$ in the distance modulus, which corresponds to a $SH0ES$ prior on $H_0$. We do not test other models (for example, adding curvature $\Omega_k$, since the luminosity distance was derived under the explicit assumption of a flat FRWL).

\section{Discussions}
The proposed modification of the luminosity distance formula is without a doubt quite significant.
The resulting matter content of the Universe is changed significantly compared to the $\Lambda$CDM model.
Now, the Universe requires  a matter with equation of state $w_{ex}\gtrsim 1$.
This seems quite strange compared to the $\Lambda$CDM model.
However, the $\Lambda$CDM model has its peculiarities  as well. 
It is called the ``concordance'' model, since it fits the observations very well, but at the cost of  pronouncing about 95\% of the energy content as ``dark'', i.e., invisible, non-interacting with light, even though the model itself is based on light observations. For now, there are no well-justified candidates for dark matter.

Besides this obvious inconsistency, there are known tensions within the model, such as the Hubble tension, which leads to the search for a proper dark energy theory. All this makes the study of the model and its alternatives even more important. In this paper, we propose, as an alternative, a cosmological model without cosmological constant, which can still fit the SN Ia observations.

Though our fit is possible for $w_{ex}\sim 0.9$, it really prefers matter with 
$w_{ex}\in [1,\;5/3]$.
Matter with $w=1$ is known as stiff matter, and we can trace its use in cosmology back to Zeldovich's seminal paper~\cite{Zeldovich1972}.
Subsequently, many other variants of the ideas in this paper have been proposed. We shall mention here Ref.~\cite{Mukherjee_2006}, where a non-linear equation of state of the form
 $p=A \rho - B\sqrt{\rho},\;\; A, \;B \in \mathbb{R}$ has been considered.
It is shown there that a matter with such an equation of state can be interpreted as a mixture of several components with linear equations of state, and one of them has
 $w=A$. 
 Due to the arbitrariness of $A$, the model can describe the matter needed in our fit.
 More examples can be found, e.g., in Ref.~\cite{PhysRevD.92.103004} and the references therein.
 
The stiff matter physical realization can be quite simple. 
For instance, a scalar field without self interaction potential has $w=1$.
For $w>1$ (ultra-stiff matter), the realization we know is within the $k-$essence model.
The $k-$essence models have a non-standard kinetic term in the form of
$L_{kin}=f_1(\varphi) f_2(X)$, where $X=(\nabla \varphi)^2/2$ and can have any equation of state~\cite{PhysRevD.63.103510}.
An example in an isotropic and homogeneous Universe is a scalar field without potential for which $f_2(X)=X^a$.
In this case, the equation of state will be 
\be 
w= \frac{f_1(\varphi)f_2(X)}{f_1(\varphi)\;2\; X\; (d f_2(X)/dX) -f_1(\varphi)f_2(X)}=
\frac{1}{2a-1},
\ee
which does not depend on $f_1$.
So, for $a \in \{1/2,1\}$, we have $w \in \{\infty, 1\}$.
A related point to consider comes from Ref.~\cite{Maartens2010}, where a mechanism for the ``stiffening'' of the equation of state is considered.
It allows us to obtain matter with $w_{ex}=1$ and $w_{ex}=5/3$ from matter with $w=0$ and $w=1/3$, respectively.
%
%
%
%

Models with ultra-stiff  matter have the intrinsic capacity  to explain the existence of a homogeneous and isotropic Universe without the need for a special inflation epoch.  The point is that the density of such matter decreases faster than the density of ordinary matter with the increase in scale factor $a(t)$.
This, in turn, means that at the beginning of the Universe, there was an epoch where ultra-stiff matter was dominant.
As discussed in our previous paper~\cite{SS2023}, in such an environment, any $U(1)$ field behaves as tachyons.
The tachyons can be considered as mirror reflections of the usual matter with respect to the speed of light---they cannot move slower than the speed of light, and their energy increases with the decrease in their speed $v$, becoming infinite at $v=c$ and tending to zero at $v \rightarrow \infty$.
Thus, they are ideal for the large-scale thermalization of the Universe.

\section{Conclusions and Outlook}
Here, we have presented how solving the 
Maxwell equations in an FLRW spacetime background
can change the basic premise of cosmology. 
By deriving the analytical solutions without relying on the geometric optics approximation and using it to find the corresponding luminosity distance, we see it differs from the standard one applied to electromagnetic observations.  
Our key finding is the possibility of a matter with equation of state $w_{ex}\geqq 1$ together with a cold dust matter to describe the data from the SN Ia. 

\begin{figure}[H]
  \includegraphics[width=0.49\textwidth]{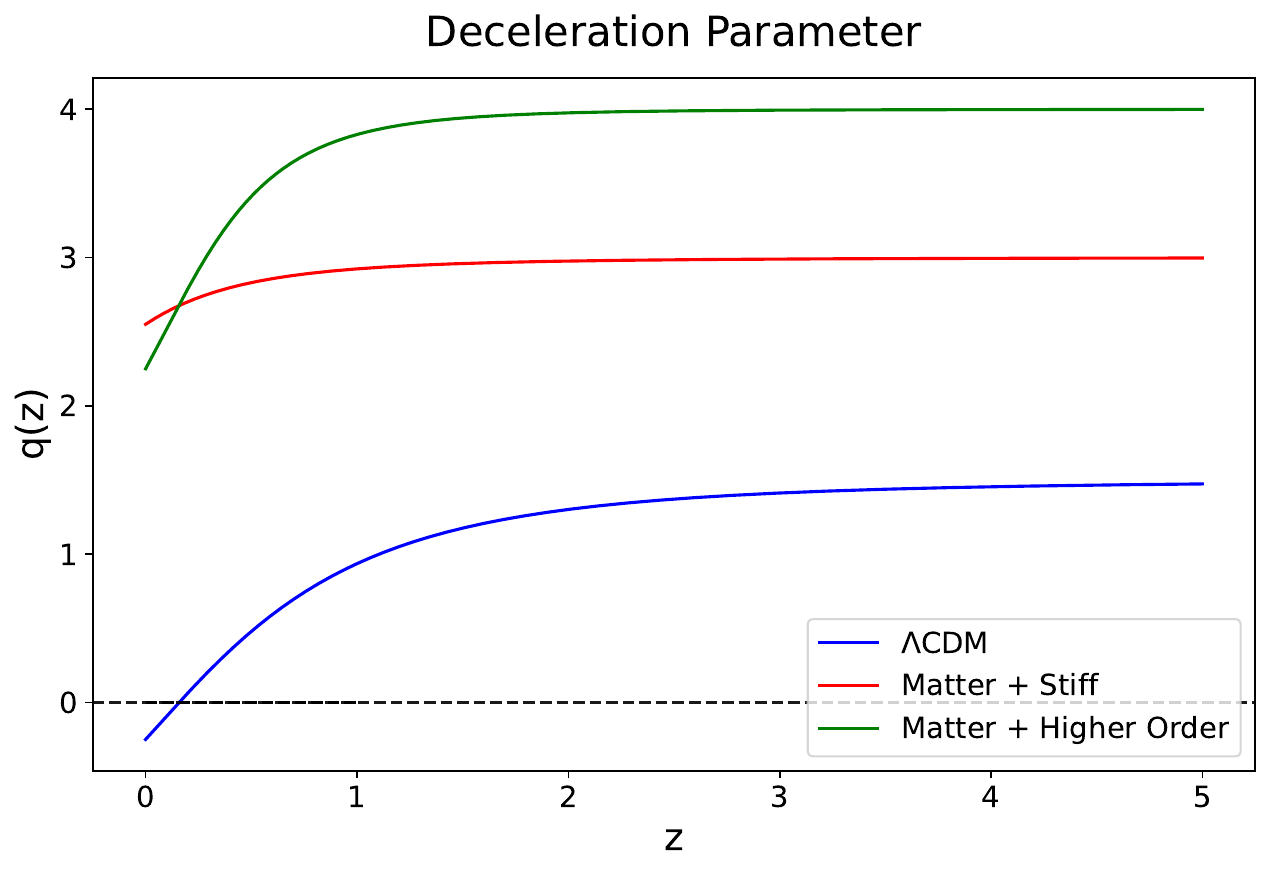}
 \includegraphics[width=0.49\textwidth]{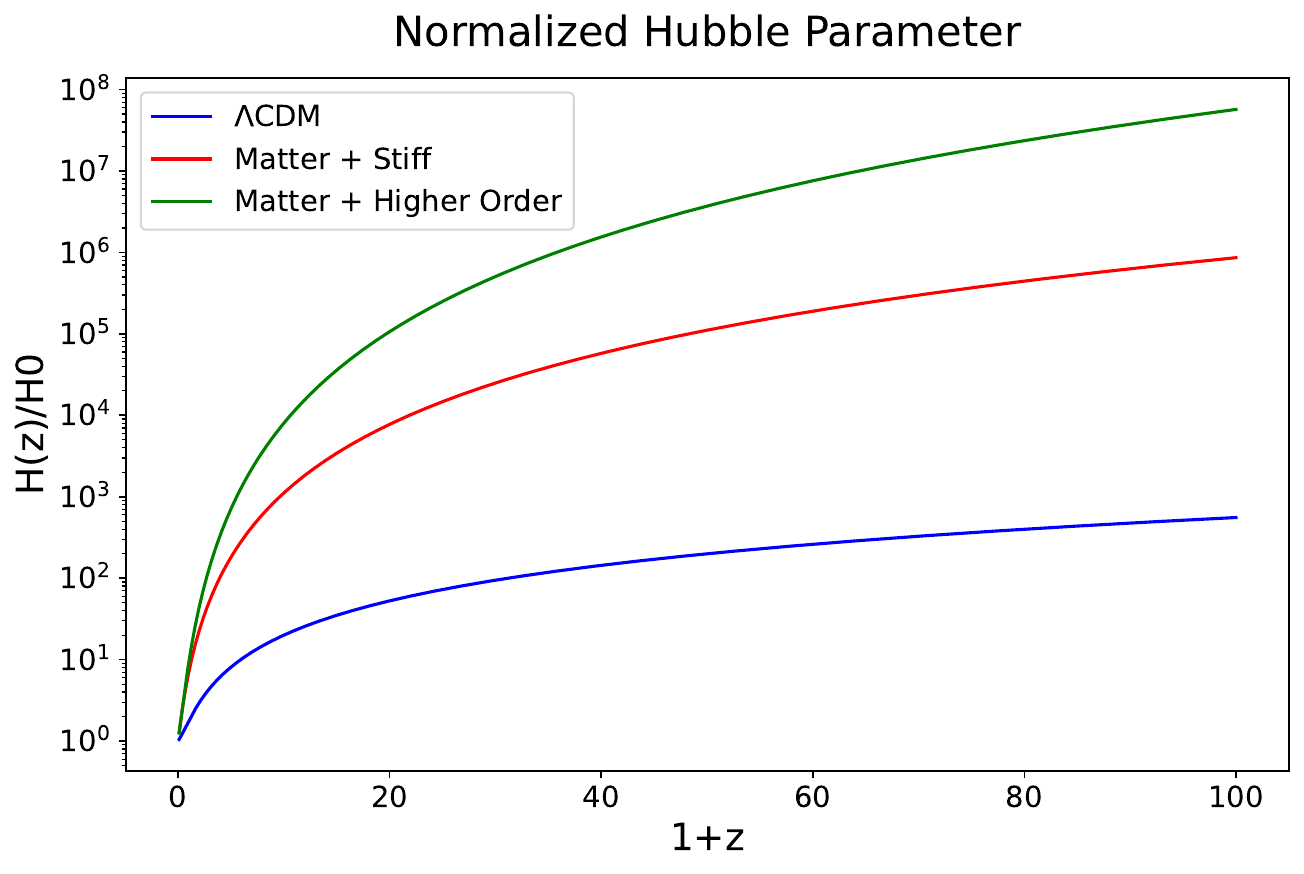}
 \caption{
 The deceleration parameter and the Hubble parameter for the parameters we obtained from fitting the data, where $H(z) = H_0 E(z)$ and $q(z) = -\frac{\ddot{a}a}{\dot{a}^2}$.}
 \label{fig}
 \end{figure}

An important result we obtained is that when utilizing the modified luminosity distance, the fits reject the cosmological constant. Therefore, there is no accelerated expansion of the Universe in this case---see Figure~\ref{fig}, where we compare the deceleration parameter and the Hubble parameter for $\Lambda$CDM and the expansions considered in this paper. We see that for the stiff-matter models, the Universe is now in a stage of  deceleration. 
While this possibility is interesting, it needs to be confirmed by adding more observational data (such as Baryonic Acoustic Oscillations and CMB). Its connection to modified gravity theories predicting an ultra-stiff term, however, makes it a promising new direction for theoretical~work. 

In conclusion, we wish to stress again the importance of thorough investigation of the propagation of light in curved spacetime.
Here, a small change in the assumptions results in a significant change in the whole picture.

\authorcontributions{Conceptualization, M.S.; methodology, D.S. and M.S.; software, D.S. and M.S.; validation,  D.S. and M.S.; formal analysis,  D.S. and M.S.; investigation,  D.S. and M.S.; resources,  D.S. and M.S.; data curation, D.S. and M.S.; writing---original draft preparation, M.S.; writing---review and editing,  D.S. and M.S.; funding acquisition,  M.S. All authors have read and agreed to the published version of the manuscript.}

\funding{This research was funded by Bulgarian National Science Fund grant number KP-06-N58/5.}

\institutionalreview{Not applicable.}

\informedconsent{Not applicable.}

\dataavailability{The original contributions presented in this study are included in the article. Further inquiries can be directed to the corresponding author.} 





\conflictsofinterest{The authors declare no conflicts of interest.} 



\abbreviations{Abbreviations}{
The following abbreviations are used in this manuscript:\\

\vspace{-6pt}
\noindent 
\begin{tabular}{@{}ll}
FLRW&Friedman--Lema\^itre--Robertson--Walker\\
$\Lambda$CDM&Cosmological constant---Cold Dust Matter\\
SH0ES&Supernova, H0, for the Equation of State of Dark Energy\\
SN Ia&Supernovae of Type Ia

\end{tabular}
}



\begin{adjustwidth}{-\extralength}{0cm}
\setenotez{list-name=Note}
\printendnotes[custom]

\reftitle{References}

\PublishersNote{}
\end{adjustwidth}
\end{document}